\documentclass[a4paper]{article}

\usepackage{amsthm}
\pdfoutput=1
\usepackage{libertine}
\usepackage[libertine]{newtxmath}
\usepackage{a4wide}

\usepackage[hidelinks,bookmarksnumbered]{hyperref}
\usepackage{cleveref}
\usepackage{xcolor}
\usepackage{xypic}
	\CompileMatrices

\usepackage[english]{babel}

\usepackage{param-reqs-macros}

\newcommand{\thetitle}{Parametricity Features and their Requirements}
\newcommand{\theauthors}{Andreas Nuyts \\ imec-DistriNet, KU Leuven, Belgium}
\begin{document}
	\addtolength{\voffset}{-.5in}

\title{\thetitle}
\date{\today}
\author{\theauthors{}}
\maketitle

\noindent%
In this note, we discuss a number of parametricity features and what their requirements are in terms of complexity of the type system and its model.

\noindent%
\emph{This note started off as \S 9.5 of my PhD thesis \cite{nuyts-phd} but required an update as I had neglected to discuss the distinction between cartesian and affine cubical models of parametricity.}

\section{Parametricity of System F}
Parametricity of (predicative \cite{predicative-system-F}) System F can be modelled using \textbf{Reynolds' original set model} \cite{reynolds}, which consists of an object interpretation and a relational interpretation relating 2 (or $n$) instances of the object interpretation. Reflexivity and the identity extension lemma are proven a posteriori by induction on the type.

\section{Parametricity of System F\texorpdfstring{$\omega$}{omega}}
Reynolds' model can be structured as a \textbf{reflexive graph} mod\-el, which allows straight\-forward generalization to a model of (predicative) System F$\omega$ \cite{fomega-parametricity}.
Reflexivity is now built into the model, and identity extension follows from the fact that type operators preserve reflexivity (and that reflexivity in the kind $*$ of types sends a node/type $T$ to the edge/relation $\EqinRel_T$).

\section{Dependently Typed Parametricity}
\subsection{Proof-Irrelevant Parametricity}
\textbf{Proof-irrelevant} parametricity (relational parametricity w.r.t.\ proof-irrelevant relations) can again be proven by a \textbf{reflexive graph} model \cite{dtt-parametricity}, which is mostly an instance of the general presheaf model of DTT \cite{Hofmann97-presheaf-chapter,psh-universes}.
Identity extension does however not hold for the universe, or for most types built using the universe.
The types for which identity extension holds, are called \textbf{discrete}. Discrete types are closed under the type operators available for small types.
So we only get \textbf{identity extension for small types}.
\index{identity!extension}%

\subsection{Proof-Relevant Parametricity}
In proof-relevant parametricity, we allow the use of proof-relevant relations.

\subsubsection{Without Identity Extension}
Atkey et al.'s universe \cite{dtt-parametricity} not only fails to be discrete, it also fails to be proof-irrelevant.
Indeed, a proof of relatedness of types $A$ and $B$ is a relation between $A$ and $B$, which is in general not unique.
So Atkey et al.'s \textbf{reflexive graph} model already covers proof-relevant parametricity without identity extension.

\subsubsection{With Identity Extension for Small Types}
This can no longer be modelled in a reflexive graph model, as it does not validate identity extension for $\Pi$-types.
Indeed, identity extension requires that two functions $f, g : (x : A) \to B(x)$ are related if and only if they are equal.
But equality implies that $f$ and $g$ have the same action on edges. In proof-irrelevant parametricity, this is automatic, because proof-irrelevance asserts that there is always at most one edge between the required source and target nodes. In proof-relevant parametricity, it is instead required that `relatedness' of $f$ and $g$ says something about their action on edges.
Thus, we do not only need edges to talk about how points are related, but also squares to talk about how edges are related. Introducing squares, creates a need for cubes to talk about how squares are related (unless we require proof-irrelevance at the square level), which requires 4-cubes, etc.
Thus, \textbf{proof-relevant} parametricity with \textbf{identity extension for small types}, should be modelled in a \textbf{cubical set} model, which is again an instance of a presheaf model of DTT.

It should be noted that these cubical sets should be \textbf{cartesian}, i.e.\ our cubes should have diagonals, modelling contraction of bridge interval variables in the type system. This is needed in order to prove discreteness of the $\Pi$-type. There are several ways to understand this:
\begin{enumerate}
	\item Discreteness of $(x:A) \to B$ means that every function $\IX \to (x:A) \to B$ from the bridge interval is constant in its first argument. In a cartesian system, this can be inherited from discreteness of $B$ by swapping both arguments. In an affine system, exchange with interval variables is only allowed in one direction so it does not constitute an isomorphism of $\Pi$-types. (Note that in the affine case we have merely given a failed proof, not a counterexample.)
	\item Using the theory of robust notions of fibrancy \cite[ch.\ 8]{nuyts-phd}: Discreteness in cartesian cubical sets is a robust notion of fibrancy \cite[ex.\ 8.4.9]{nuyts-phd}, implying that discretenss of the $\Pi$-type is inherited from the codomain. For affine cubes (without diagonals) we get no such result: the pullback of $\IX \to \top$ along $\IX \to \top$ is $\IX \times \IX \to \IX$; the cartesian product contains a square and an \emph{unassociated} diagonal. Requiring this to be a left map would entail that all discrete types are also proof-irrelevant. (Again, this is merely a failed proof of robustness, not a counterexample to discreteness of the $\Pi$-type.)
	\item In an affine cubical setting, if the type $A$ contains a non-trivial bridge (which is even possible if $A$ is discrete: discrete types can still contain non-trivial heterogeneous bridges) then a function $\Gamma \sez f : \IX \to (x:A) \to B$ introduces an a priori non-trivial square in $B$ with an unassociated diagonal. If $B$ is discrete, the square will be flat, but we cannot reassociate the diagonal. (This is a counterexample.)
\end{enumerate}
\noindent%
The depth 0 mode of Degrees of Relatedness \cite{reldtt,reldtt-techreport} is a type system for proof-relevant parametricity with identity extension for small types\footnote{Arguably for all types, as it is impossible to mention the universe of depth 0 types, which is a depth 1 type.}, modelled exactly in the category of cartesian cubical sets.

\subsection{Internal/iterated parametricity}
In type systems with internal parametricity, there is typically a type $x \bridge_T y$ of proofs (called bridges) that $x$ and $y$ are related.
Typically, the internal parametricity operators can be applied to the above type (unless the type system takes special measures to prevent it, such as including a pointwise modality in ParamDTT \cite{paramdtt}, which we consider a flaw and not a feature and which we rectified in RelDTT \cite{reldtt}), allowing internal reasoning about relatedness of proofs of relatedness.
We call this iterated parametricity.

A first observation is that the index $T$ in $x \bridge_T y$ can itself be a bridge type, yielding a square type. If $T$ is a square type, we end up with a cube type, and so on.
The fact that we can mention squares and higher cubes in the syntax clearly requires an (affine, cartesian or other) \textbf{cubical set} model.

Other than that, internal or iterated parametricity does not add much to the concerns already mentioned: the type $x \bridge_T y$ brings the same challenges as the type $(x : A) \to B(x)$ when $A$ has a non-trivial edge.
As such, internal/iterated parametricity requires a \textbf{cartesian cubical model} if you also want \textbf{proof-relevance and identity extension for small types}, but not otherwise.

\subsubsection{BCM-style operators}
BCM's operators $\Phi$ (a.k.a.\ $\name{extent}$) and $\Psi$ (a.k.a.\ $\name{Gel}$) \cite{moulin} for relating functions and types respectively, require \textbf{affine cubical sets}. This is because nested use of these operators simply does not specify any behaviour on diagonals.
We generalized $\Psi$ to also work for \textbf{non-affine base categories} \cite{transpension}, but there are still many questions surrounding this approach, especially when it comes to implementation.

\subsubsection{\texorpdfstring{$\Glue$}{Glue} and \texorpdfstring{$\Weld$}{Weld}}
When opting for a cartesian cubical set model instead -- e.g.\ because identity extension is desired -- one can instead use the $\Glue$ and $\Weld$ operations \cite{cubical,paramdtt}\cite[\S 6.3]{nuyts-phd}. However, these do not give full control over the relational content of bridges in the universe, nor over the relational action of bridges between functions.

\subsection{Identity extension for large types}
\index{identity!extension}%
If we want to attain identity extension for large types in DTT, a deep intervention in how the type system works is appropriate.
In order to see why, we consider a simple example. In System F, the type $\forall X.B$ (with $B$ closed) only contains constant functions. This is a parametricity result.
In DTT, this type translates to $(X : \uni{\ell}) \to B$.
Now if we instantiate $B := \uni{\ell}$, then we can inhabit the type with the identity function $\lambda(X: \uni{}).X$, which is not constant.

There are several perspectives to understand this problem.
A first perspective is by observing that the $\Pi$-types of DTT generalize both the parametric $\forall$-type former from System F (which forbids its inhabiting functions to inspect their argument) and the non-parametric type/kind former $\to$ (which allows its inhabiting functions to inspect their arguments).
As such, the $\Pi$-type has non-parametric inhabitants coming from the arrow type, violating the parametricity results that held for the $\forall$-type.
In ParamDTT \cite{paramdtt}, we rectified this by reintroducing a parametric function type $(\ctxmod{}{\parmod}{x}{A}) \to B(x)$ containing only parametric functions.

Alternatively, we may consider types' levels. Concretely, we expect inhabitants of $(X : \uni \ell) \to B$ (with $B$ closed) to be constant \emph{assuming that $B$ has level $\ell$}, thus ruling out $B = \uni \ell$ which has level $\ell+1$.
This would follow from a generalization of Atkey et al.'s result \cite{dtt-parametricity}, which asserts identity extension for small types.
However, if we inspect their model, it turns out that smallness is not really the property they rely on; rather, they rely on the fact that \emph{discrete} types satisfy identity extension and that discrete types are closed under small type formers.
It is to be expected that future work on parametricity, which may feature relational higher inductive types (see the HoTT-book for HITs \cite{hottbook}), will provide small type formers which do not preserve discreteness.
Thus, it seems wise to distinguish a type's level (a stratification useful for guaranteeing predicativity) from its \emph{depth} \cite{reldtt} (i.e. its relational complexity).
\index{depth}%

RelDTT \cite{reldtt} cleanly unifies these two perspectives: the function type $(\ctxmod{}{\mu}{x}{A}) \to B(x)$ gets annotated with a modality $\mu$. The choice of possible modalities $\ismod \mu m n$ depends on the depths $m$ of $A$ and $n$ of $B$. If $A$ and $B$ have the same depth, then continuity $\ismod \conmod m m$ becomes an option and we can consider functions that inspect their argument. If $B$ has depth $n = m-1$, then we can form the type $(\ctxmod{}{\parmod}{x}{A}) \to B(x)$ whose elements satisfy free theorems \cite{theorems-for-free}, but thanks to the wealth of modalities in RelDTT we can still consider ad hoc polymorphic functions $(\ctxmod{}{\hocmod}{x}{A}) \to B(x)$ allowing e.g.\ postulation of the law of excluded middle $(\ctxmod{}{\hocmod}{X}{\uni{}}) \to X \uplus (X \to \Empty)$ without breaking parametricity in general.

In summary, if we care for dependently typed parametricity with \textbf{identity extension even for large types}, then we should be looking towards modalities or at least a stratification of types based on their relational complexity (which may or may not be decoupled from the level).
I would argue that RelDTT is so general that such modal or stratified systems should be almost always explicable as a subsystem of RelDTT, e.g.:
\begin{theorem}
	There exists a non-trivial model of DTT with Agda-style cumulativity\footnote{i.e.\ if $X : \uni \ell$ then $\name{Lift}\,X : \uni{\ell + 1}$ and $\name{Lift}\,X \cong X$.}, in which any function $f : \uni \ell \to A$ where $A : \uni \ell$, is constant. \cite{nuyts-phd}
\end{theorem}
\begin{proof}
	DTT translates to a subsystem of RelDTT where types of level $\ell$ automatically also have depth $\ell$ and all functions are mediated by $\ismod{\parmod^k}{m + k}{m}$ or $\ismod{\strmod^k}{m}{m+k}$ or $\ismod{\conmod}{m}{m}$, and where $\name{Lift}\,X := \Modifynovar{\strmod}{X}$.
	
	Under this translation, $f$ ends up having type $(\tymod{}{\parmod}{\uni \ell}) \to A$. Now we can build a 1-edge between any two types in $\uni \ell$ (e.g. by welding over the empty type), yielding a 0-edge in $A$, which is constant by the degeneracy axiom of RelDTT \cite{reldtt}.
\index{axiom!degeneracy}%
\end{proof}

I hope that this exposition makes it clear that, \emph{in general}, dependently typed parametricity does \emph{not} require modalities, even if we require identity extension for small types.

\bibliographystyle{alphaurl}
\bibliography{param-reqs-refs.bib}

\end{document}